# Compiling Chemical Knowledge into Executable Descriptors for Materials Prediction

Jaehwan Choi[1], Kunik Jang[1], Seongmin Kim[1], Shuan Chen[1], Kyungju Nam[3], Seung Hyo Noh[3], Donghwi Kim[3], and Yousung Jung[1,2*]

[1] Department of Chemical and Biological Engineering (BK21 four), and Institute of Chemical Processes, Seoul National University, 1 Gwanak-ro, Gwanak-gu, Seoul 08826, Korea

[2] Institute of Engineering Research, Seoul National University, 1 Gwanak-ro, Gwanak-gu, Seoul 08826, Korea

[3] Materials Research & Engineering Center, Hyundai Motor Company, Uiwang-si, Gyeonggi-do, Republic of Korea

* Email: yousung.jung@snu.ac.kr

## ABSTRACT

Materials prediction depends critically on how scientific knowledge is represented, yet many governing considerations exist only as natural-language heuristics that conventional learners cannot use. We introduce CRISP, a large language model-assisted framework that treats representation construction as a rule-space exploration and compilation problem: it repeatedly samples target-relevant chemical rules without access to structures, labels or data splits, consolidates related concepts, and compiles each into an executable scalar descriptor supplied to a conventional learner. For positive–unlabeled inorganic-crystal synthesizability, CRISP outperformed expert-curated and generic structural representations under a shared learner and surpassed purpose-built synthesizability models, with its advantage most pronounced under structural-size and chemical-family shifts. Infrequently generated rules contributed complementary predictive information, showing that generation frequency does not determine utility. The same workflow yielded competitive representations for formation energy and ionic conductivity while revealing task-dependent limits for shear modulus, establishing a dataset-blind, auditable route from broad chemical knowledge to transferable computational representations.

Materials prediction is often limited less by the choice of learner than by how the target is represented[1]. For example, formation energy is a defined number that a calculation returns for any given structure. Synthesizability is not: whether a predicted crystal can be made depends on where it sits relative to the convex hull, on whether a kinetically accessible route exists, on which precursors are at hand, on conditions that no single computed quantity summarizes, and on whether anyone has tried[2–4]. Each of those considerations is real, and each is normally stated as a natural-language sentence rather than as a number. Conventional learners cannot read sentences. Deciding which considerations become variables is therefore part of the scientific problem, and it has to be settled before anything can be learned.

Two responses are common. The first is to let a model infer its own variables, from a crystal graph or from a text rendering of the structure[5–8], and recent work has pushed large language models (LLMs) into the predictor role itself[9–13]. Accuracy can be high, but the variables that drive a prediction stay out of view, and a rationale generated alongside the answer is produced after the fact and need not describe what the model actually used[14–16]. The second is to hand-engineer descriptors, which are transparent and computable, but a fixed catalog holds only the concepts its designers thought to include[17,18].

Prior work has shown that LLM-generated knowledge can be transformed into interpretable features for molecular property prediction, combining established information extracted from scientific text with patterns inferred from labelled molecular data[19]. Here we formulate a distinct methodological problem: how to systematically explore the broad, stochastic space of scientific rules accessible to an LLM and compile that distributed knowledge into an executable representation. This problem becomes particularly demanding for periodic crystals, where coordination environment, packing, symmetry, phase competition, and extended connectivity require nontrivial calculations over three-dimensional structures to be expressed as explicit descriptors[18,20]. Moreover, because any individual LLM response captures only a limited sample of this rule space, a representation constructed from a small number of responses may depend strongly on prompt and sampling variability.

We address these challenges with CRISP, an LLM-assisted workflow that repeatedly samples target-relevant chemical rules without access to structures, labels, target values or data splits, consolidates semantically related concepts, and compiles each concept into an auditable scalar program operating on a periodic crystal structure. The resulting catalog is frozen and supplied to a conventional learner, with the LLM taking no part in prediction. Every variable therefore corresponds to a named chemical rule with an inspectable program behind it, allowing individual descriptors to be traced, ablated or rewritten[21,22]. CRISP thus converts pretrained chemical knowledge into a dataset-blind computational prior while making rule-space coverage an experimental variable. This design allows us to test whether systematic exploration beyond commonly articulated heuristics yields complementary predictive information under distribution shift.

We take inorganic-crystal synthesizability as the primary test, because a useful representation has to carry thermodynamic, structural, compositional, and coordination information at once[2,3], and much of what matters has existed only as a verbal heuristic rather than as anything a standard descriptor set precomputes. In this task a reported structure is a positive and an unreported one is not a negative, so the setting is positive–unlabeled[5,6,11,12,23–26]. We make two comparisons. The first holds the learner and the entire evaluation protocol fixed and changes only the descriptors, spanning local-coordination, distribution-based, and structure-text representations together with an expert-curated set[18,27,28]. The second compares CRISP against purpose-built synthesizability models, including fine-tuned LLM predictors and a crystal-graph network, which differ in predictor as well as representation; CRISP outperforms all of them[5,9,11–13].

The same workflow is then aimed at three regression targets that differ in scale and in what governs them: formation energy, shear modulus, and room-temperature ionic conductivity[29], the last of these small enough to ask whether a compact catalog is worth building at all. The regression results line up with the synthesizability one: the advantage appears wherever the test cohort leaves the development distribution or the data run short. The concepts a chemist would think to formalize are not the same as the concepts that carry a prediction, and it is the difference between those two sets that transfer exposes. Chemical intuition remains the source of the rules; what it no longer has to be is the filter on which of them a model is allowed to use.

# Results

## Exploring and compiling chemical rules into executable crystal descriptors

We constructed CRISP as a five-step workflow that separates chemical-knowledge compilation from statistical prediction (**Figure 1**). An LLM first explores target-relevant chemical rules (**Figure 1a**), semantically related rules are consolidated (**Figure 1b**), and each consolidated rule is compiled into an executable scalar function (**Figure 1c**). Applying these functions to a crystal produces a frozen descriptor vector (**Figure 1d**), which a conventional learner uses for prediction and feature-level interpretation (**Figure 1e**).

For synthesizability, *gpt-4.1-mini* generated ten candidate rules per response across 1,000 independent calls, yielding 10,000 rules. We selected a non-reasoning model to favor rapid, divergent exploration rather than convergence on a single answer. The resulting pool contained both recurring principles, such as thermodynamic stability and coordination, and less frequently articulated structural heuristics. We grouped semantically related rules and consolidated them into 50 representative families spanning thermodynamic, compositional, geometric, coordination, symmetry, and other structure-sensitive considerations (**Figure S1**). Coverage and convergence analyses support 1,000 calls as a practical saturation budget (**Figure S2**), and all 50 rules are listed in **Table S1**.

Each consolidated rule was then compiled into an explicit scalar function operating on crystal-structure and associated physicochemical inputs. Pretrained synthesizability models and learned surrogates were excluded, limiting the transfer of task-specific teacher-model bias and keeping every descriptor traceable to a named chemical rule and an inspectable calculation. No human editing was applied during rule consolidation. Human intervention during code preparation was limited to corrections required for execution, without altering the chemical logic or computational structure proposed by the LLM.

The resulting 50-dimensional descriptor catalog was frozen before model fitting. A bagged PU Random Forest was selected as the downstream consumer because it combined competitive performance among the tested conventional learners with stable ensemble predictions and direct feature attribution (**Figure S3**). The regression applications retain the same separation between the compiled representation and the conventional learner.

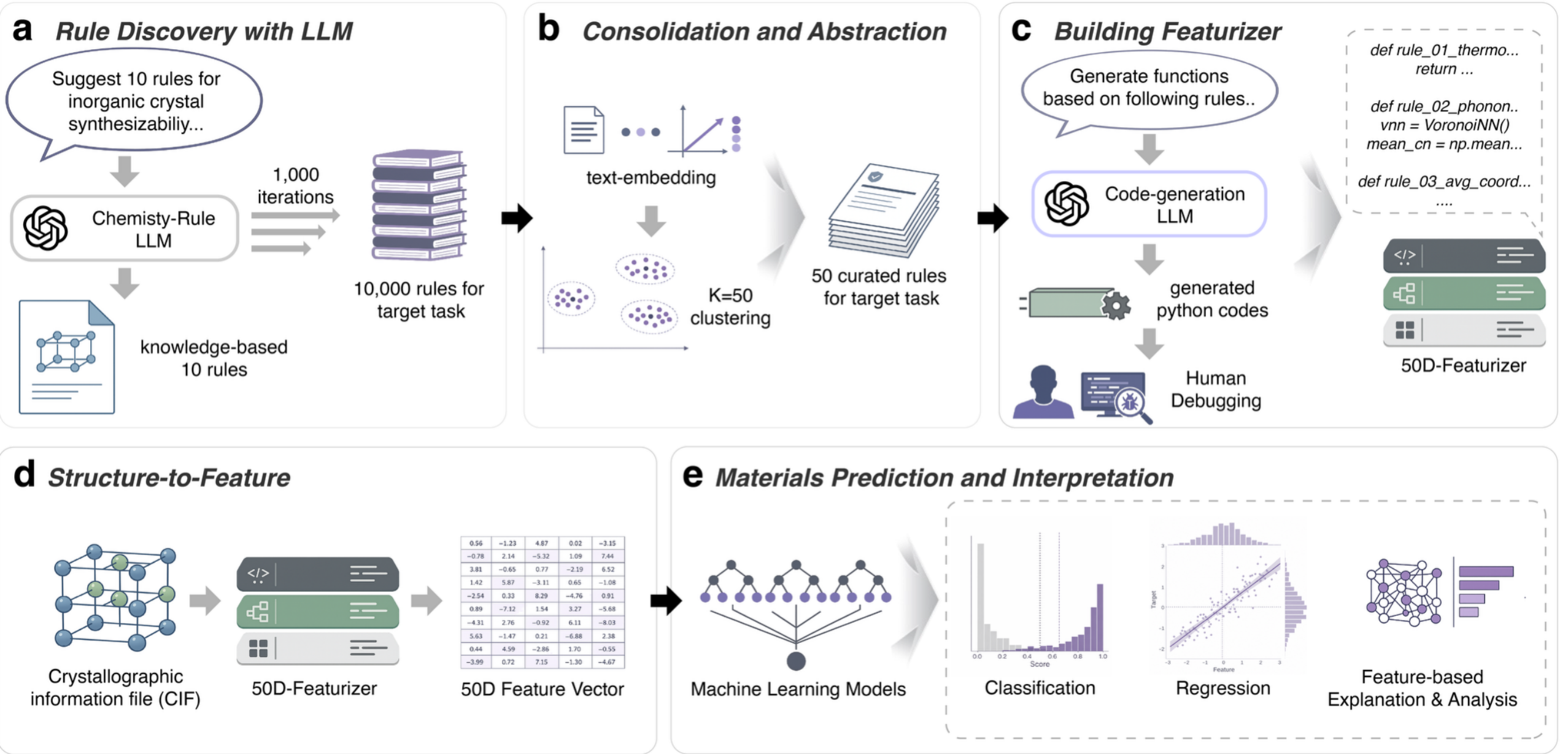

**Figure 1. CRISP workflow for compiling chemical rules into executable crystal descriptors.** The panels show the synthesizability instantiation; the same five steps are used for every target, with the objective, permitted inputs, catalog size, and learner specified per task. Panel values correspond to the synthesizability catalog (K = 50); the regression catalogs use K = 25. (a) A target-specific prompt samples chemical and structural rules from an LLM. (b) Related rule texts are embedded, organized, and consolidated into representative rules. (c) Each rule is compiled into an executable scalar descriptor and checked for code execution and prohibited operations. (d) The frozen catalog maps the permitted material inputs to a compact task-specific vector. (e) A conventional statistical learner consumes the vector for materials prediction; for synthesizability this is a bagged random forest.

## Synthesizability as a test of compiled concept coverage

We first evaluated CRISP on inorganic-crystal synthesizability because this target combines several chemical and structural signals, none of which is sufficient on its own. We used a standardized positive–unlabeled (PU) benchmark in which experimentally reported structures are labeled positive and hypothetical or not-yet-reported structures remain unlabeled. To avoid differences arising from model-specific datasets, all models were trained and evaluated on the same MP30 split constructed from the Materials Project database[26]. The benchmark contains 100,194 structures with at most 30 atoms per unit cell, including 38,347 positive and 61,847 unlabeled structures, with 10% of each group reserved as a frozen test set. This common benchmark separates methodological differences from variation in the underlying positive and unlabeled pools.

We first isolated the effect of representation by holding the complete PU-RF training and evaluation protocol fixed. Only the mapping from a crystal structure to a numerical vector was changed. The six representations were CRISP, an expert-curated set, CrystalNN, CFID, a raw-CIF embedding, and a Robocrystallographer-text embedding. Performance differences in this experiment therefore reflect the utility of each featurization under the same downstream learner. Because the representations use different permitted inputs, the comparison measures complete featurization-pipeline utility rather than strict information parity.

We then compared the complete CRISP pipeline with PU-CGCNN, CSLLM, SynCry, and StructLLM. We reconstructed these published model architectures and trained or fine-tuned each one on the same frozen MP30 training identifiers before evaluation on the identical held-out test set. These results compare complete prediction pipelines, since both the representation and predictor vary across models. Prompt-only *GPT-4o* and *Llama-3-8B* evaluations were also included to provide unadapted LLM baselines. **Table 1** summarizes the representations and model pipelines used in both comparisons.

**Table 1. Synthesizability prediction on the MP30 benchmark.**

| Model/representation | Methodology | Input representation | TPR* | PREC* | F1* |
|---|---|---|---|---|---|
| CRISP (this work) | LLM-compiled descriptors + Random Forest | 50 executable chemistry-rule descriptors | **0.855** | **0.858** | **0.856** |
| Other representations under the identical PU-RF | | | | | |
| Expert-curated | Random Forest | 18 expert-selected composition and crystal-structure features | 0.818 | 0.850 | 0.834 |
| CrystalNN[18] | Random Forest | 122 local-coordination fingerprints | 0.665 | 0.666 | 0.665 |
| CFID[27] | Random Forest | 1,557 chemistry, cell, radial, and angular descriptors | 0.597 | 0.584 | 0.590 |
| Robocrystallographer[28] | Random Forest | 256-dimensional embedding of a generated structure description | 0.458 | 0.421 | 0.438 |
| Raw CIF | Random Forest | 256-dimensional embedding of the CIF text | 0.388 | 0.389 | 0.388 |
| Published bespoke models and general LLMs | | | | | |
| CSLLM[13] | Fine-tuned LLM | Material string with space-group information | 0.802 | 0.786 | 0.794 |
| StructLLM[12] | Fine-tuned LLM | Robocrystallographer-generated structure description | 0.728 | 0.728 | 0.728 |
| SynCry[9] | Fine-tuned LLM | Material string without space-group information | 0.707 | 0.708 | 0.707 |
| PU-CGCNN[5] | Graph-based PU learning | Crystal graph | 0.596 | 0.598 | 0.597 |
| GPT-4o | Prompted general-purpose LLM | Material string without space-group information | 0.073 | 0.073 | 0.073 |
| Llama-3-8B | Prompted general-purpose LLM | Material string without space-group information | 0.076 | 0.072 | 0.074 |

*Metrics Calibrated using the estimated $\alpha$.

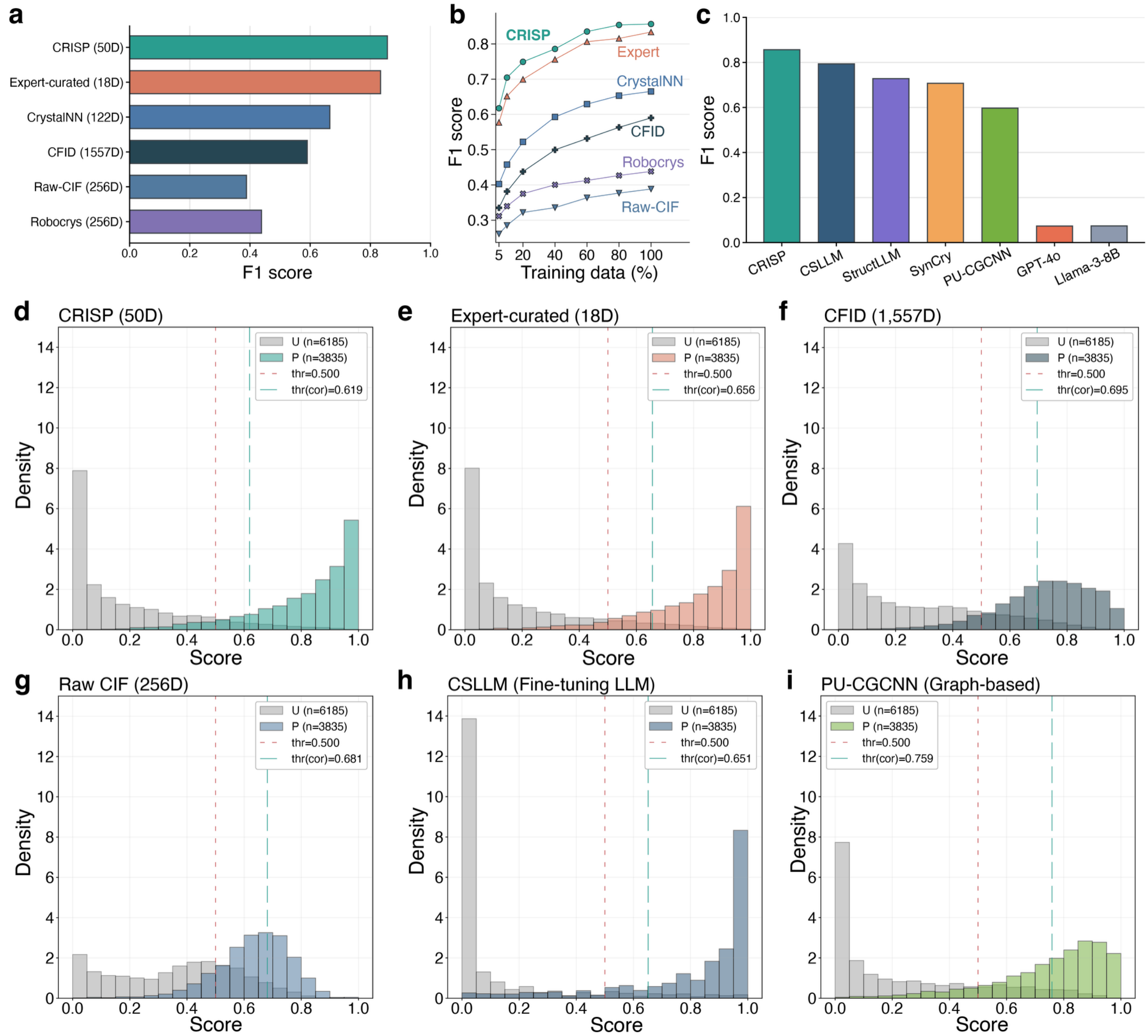

**Figure 2. Synthesizability implementation of CRISP.** (a) Estimated α-calibrated F1 score at the α=0.0567 for six featurization pipelines evaluated with the identical PU random-forest consumer, MP30 identifiers, outer split, bagging scheme, threshold calibration, and metric definitions; only the pipeline mapping a material to a vector differs. Descriptor dimension is given in parentheses. (b) The same six pipelines under nested reductions of the shared training partition. Every representation receives the same nested subsets and the same held-out structures at each fraction. (c) Comparison with published end-to-end synthesizability models and with general-purpose LLMs prompted without task adaptation. (d–i) Held-out score distributions for known positives (P, n = 3,835) and unlabeled structures (U, n = 6,185), shown for representations evaluated with the fixed random forest (d–g) and for two bespoke models (h, i). Red dashed lines mark the nominal threshold of 0.5 and teal dashed lines the α-calibrated threshold. All panels use the identical held-out MP30 test set of 10,020 structures.

**Figure 2a** summarizes the matched representation comparison. CRISP achieved the highest F1 score, with the expert-curated set as the only comparable representation. Performance did not follow vector size: CRISP (50D) and the expert-curated control (18D) were the strongest, whereas the 1,557-dimensional CFID representation performed substantially below both under the same learner. The same ordering for TPR shows that this ranking was not solely due to the PU precision correction. At 5% of the training partition, CRISP retained an F1 of 0.618, compared with 0.577 for the expert-curated set and 0.261–0.403 for the four generic representations (**Figure 2b**). CRISP also exceeded the strongest published task-specific predictor by approximately 0.06 in F1 (**Figure 2c**). Performance increased through K = 50 but plateaued at K = 60, where semantic overlap increased, supporting K = 50 as a balance between predictive coverage and redundancy (**Figure S4**).

The score distributions further clarify these differences (**Figure 2d–i**). CRISP and the expert-curated representation both concentrated known positives at high scores and much of the unlabeled set at lower scores, although CRISP achieved the higher F1. CFID and the raw-CIF embedding showed greater overlap between the two groups. Among the bespoke models, CSLLM produced an almost binary distribution with limited resolution within either score mode, whereas PU-CGCNN assigned a broad upper-score tail to unlabeled structures. CRISP instead retained a continuous score suitable for candidate ranking as well as classification.

CRISP and the expert-curated representation also showed the highest concordance in held-out score rankings (**Figure S5**), yet 339 known positives were recovered only by CRISP. The compiled catalog therefore preserves much of the expert-informed ranking while recovering a complementary subset that may reflect considerations underrepresented in manual descriptor design. No individual descriptor showed near-deterministic class separation (**Figure S6**), indicating that CRISP combines multiple partially informative chemical and structural signals.

## Retention of predictive signal under structural, chemical, and temporal shifts

High benchmark performance can reflect proximity to the development distribution and does not establish whether a representation captures transferable chemical information. We therefore evaluated CRISP under two controlled transfer settings and one retrospective temporal checkpoint. The controlled experiments applied MP30-trained models to structures with larger stored unit cells (MP > 30) and withheld entire chemical families from model development. These shifts test whether descriptors that are partly redundant within MP30 provide complementary information when structural or chemical coverage changes. The temporal checkpoint separately examined whether CRISP scores anticipated structures recognized as positive in a later database snapshot.

Across the six representations examined in Figure 2, CRISP retained the strongest performance under both structural-size and chemical-family shifts (**Figure 3a–c**). In the MP > 30 setting, CRISP achieved an F1 of 0.420, followed by the expert-curated representation at 0.384 and CFID at 0.355; the remaining representations ranged from 0.278 to 0.318 (**Figure 3a**). The separation widened when entire chemical families were withheld. CRISP reached a macro-averaged F1 of 0.537, compared with 0.373 for the expert-curated representation, 0.344 for CrystalNN, and 0.140–0.235 for the remaining representations (**Figure 3b**). It exceeded the expert-curated control in all eight held-out families (two-sided exact sign test, $p = 0.0078$; Figure 3c and Table S2). The difference was especially pronounced for intermetallics, where CRISP achieved an F1 of 0.764, compared with 0.508 for CrystalNN and 0.418 for the expert-curated representation. The consistent family-level advantage indicates that the compiled catalog retains family-relevant chemical information beyond that represented by a compact human-selected set.

We next compared CRISP with representative published text- and graph-based synthesizability models. Under the structural-size shift, CRISP retained a calibrated TPR of 0.839, compared with 0.754 for CSLLM and 0.574 for PU-CGCNN (**Figure 3d**). Across the held-out chemical families, CRISP achieved higher macro-averaged TPR and precision (0.587 and 0.589) than CSLLM (0.483 and 0.477) and PU-CGCNN (0.244 and 0.243; **Figure 3e**). The transfer advantage therefore appeared both in the controlled representation comparison and against complete end-to-end prediction pipelines.

Finally, the retrospective temporal checkpoint examined structures that were unlabeled in the 2022 database snapshot but appeared as positive entries in 2026 (**Figure 3f**). Their CRISP scores were concentrated in the high-score region, with 94.6% exceeding the nominal threshold of 0.5.

Together, the controlled transfer experiments and temporal checkpoint show that the principal empirical advantage of CRISP is not confined to a small increase in in-distribution F1. The compiled representation retains predictive signal under structural-size, chemical-family, and temporal shifts.

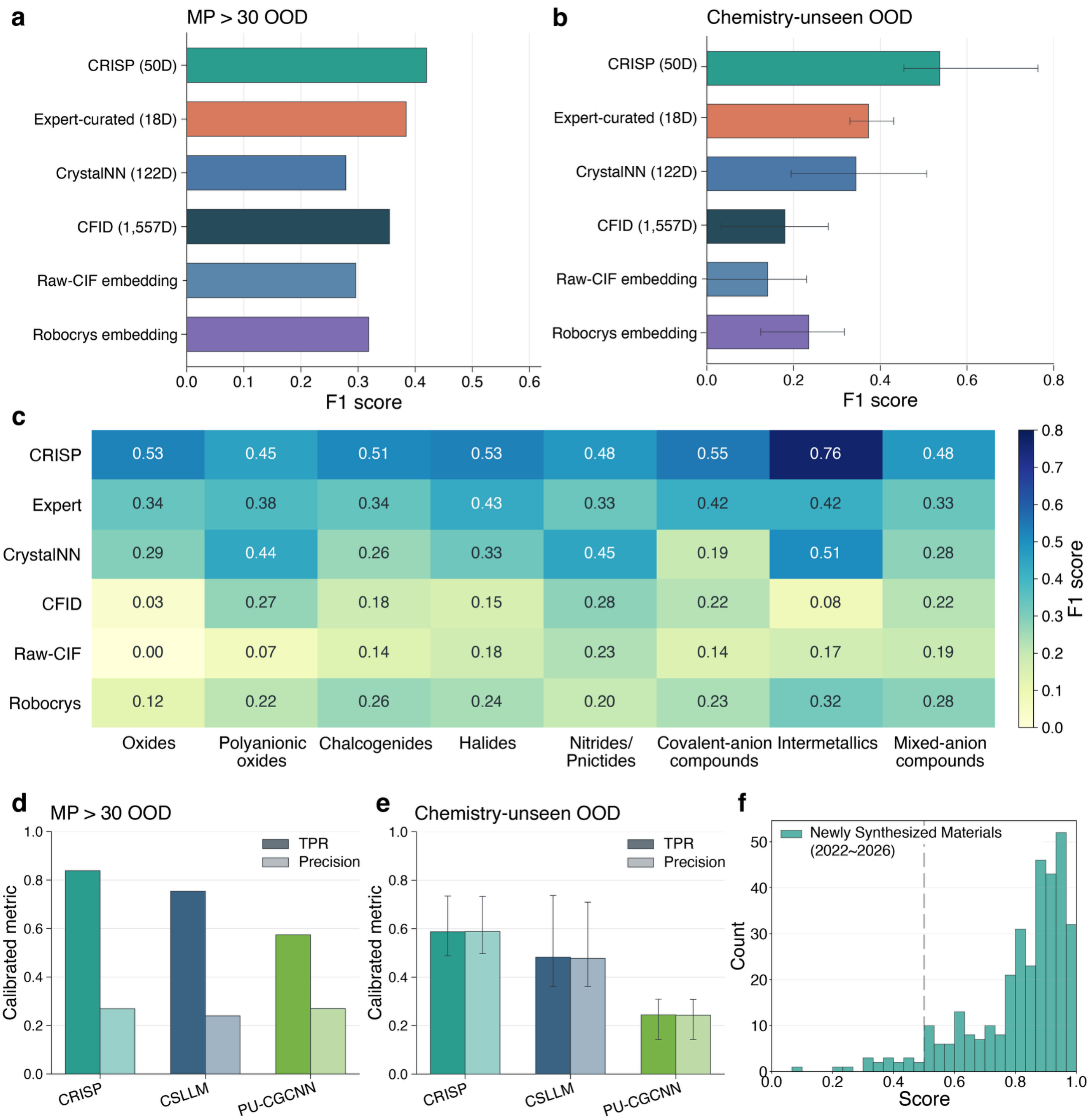

| | Oxides | Polyanionic oxides | Chalcogenides | Halides | Nitrides/ Pnictides | Covalent-anion compounds | Intermetallics | Mixed-anion compounds |
|---|---|---|---|---|---|---|---|---|
| CRISP | 0.53 | 0.45 | 0.51 | 0.53 | 0.48 | 0.55 | 0.76 | 0.48 |
| Expert | 0.34 | 0.38 | 0.34 | 0.43 | 0.33 | 0.42 | 0.42 | 0.33 |
| CrystalNN | 0.29 | 0.44 | 0.26 | 0.33 | 0.45 | 0.19 | 0.51 | 0.28 |
| CFID | 0.03 | 0.27 | 0.18 | 0.15 | 0.28 | 0.22 | 0.08 | 0.22 |
| Raw-CIF | 0.00 | 0.07 | 0.14 | 0.18 | 0.23 | 0.14 | 0.17 | 0.19 |
| Robocrys | 0.12 | 0.22 | 0.26 | 0.24 | 0.20 | 0.23 | 0.32 | 0.28 |

**Figure 3. Retention of synthesizability-prediction signal under structural, chemical, and temporal distribution shifts.** (a) Representation-controlled transfer from MP30 to 23,525 Materials Project structures containing 31–49 atoms per unit cell. Bars show the F1 score obtained using the same PU-RF consumer and thresholds calibrated on the MP30 training data. (b) Macro-averaged F1 score across eight chemistry-unseen evaluations, each conducted by excluding one composition-defined material family during training. Whiskers indicate the minimum and maximum values across the eight held-out families and therefore describe between-family variation rather than uncertainty in the macro average. (c) Family-resolved F1 score for the six representations in the chemistry-unseen evaluation. (d) Calibrated TPR and estimated PU precision of the original CRISP, CSLLM, and PU-CGCNN pipelines on the MP > 30 site-count-shift setting. (e) Macro-averaged calibrated TPR and estimated PU precision of the three model-specific pipelines across the eight held-out material families. Whiskers show the corresponding minimum-to-maximum range across families. (f) Retrospective CRISP-score distribution for 334 structures that were unlabeled in the 2022 database snapshot and appeared as positive entries in the 2026 snapshot. Here, CRISP was trained on structures using data from the 2022 database.

## Chemical interpretation and physical audit of CRISP predictions

CRISP uses an LLM to compile chemical knowledge into executable descriptors and a Random Forest to predict from them. This separation permits audits of individual rules, physicochemical families, and independently computed structural responses.

We first examined the global descriptor-importance profile (**Figure 4a**). Feature indices follow LLM cluster-frequency order; lower indices denote more frequently generated concepts. Energy above hull was the leading descriptor, but generation frequency was not significantly associated with either Random Forest importance or mean absolute SHAP attribution (**Figure S7**). Less frequently generated rules, including stoichiometry deviation (R42) and the antisite formation-energy proxy (R45), contributed substantially. Generation frequency reflects how readily a concept emerges, whereas predictive importance depends on the utility of its executable implementation. Repeated fits, held-out permutation importance, and SHAP analysis consistently identified the leading descriptors (**Figure S8**), although small rank differences among correlated or weak features were interpreted cautiously.

Using training-derived rankings, we retrained CRISP with progressively larger top- and bottom-ranked descriptor subsets (**Figure 4b**). The top ten descriptors achieved an F1 score of 0.816, whereas bottom-ranked subsets recovered more slowly, indicating that a compact core carries much of the in-distribution signal.

Post hoc expert annotations nevertheless showed that predictive information was distributed across thermodynamic, graph-relational, geometric, and compositional families (**Figure 4c, d** and **Figure S9**). Geometry plus composition nearly matched the full catalog in distribution, but the full catalog performed better under both structural-size and chemical-family shifts in a separate within-pipeline ablation (**Figure S10**). Thus, descriptors that appear redundant within MP30 can become informative after distribution shift.

Descriptor importance also varied across thermodynamic regimes. Energy above hull lost importance on the convex hull, while defect-tolerance, cell-complexity, and stoichiometric descriptors became more prominent (**Figure S11**). Taxonomy-resolved SHAP analysis further identified shared global signals and family-specific contributions (**Figure S12**).

Finally, we tested whether CRISP scores were associated with DFT responses to local structural perturbations (**Figure 4e–g**). We analyzed 50 high-scoring known positives, 50 high-scoring unlabeled structures, and 50 low-scoring unlabeled structures. For each parent, ten structures generated by 1% random coordinate displacements were relaxed under the same DFT protocol (**Figure S13**). We counted perturbations relaxing at least 0.02 eV/atom below each parent; higher counts indicate lower local structural robustness.

The high-score groups had fewer energy-lowering perturbations than the low-score unlabeled group (**Figure 4e–g**). All ten perturbations crossed the threshold for 1 of 50 high-scoring known positives, 4 of 50 high-scoring unlabeled structures, and 12 of 50 low-scoring unlabeled structures. Representative parents and relaxed perturbations are shown in **Figure S14**. Thresholds of 0.01 and 0.03 eV/atom reproduced the same qualitative ordering (**Figure S15**). Among parents for which all ten perturbations reached lower energies, the high-score groups remained concentrated at small parent-relative CrystalNN-fingerprint distances, whereas the low-score unlabeled group showed broader structural changes (**Figure S16**).

Together, the feature-level and DFT analyses show that CRISP combines distributed chemical information into scores associated with local structural robustness. This provides an independent physical consistency check without treating relaxation behavior as direct evidence of synthesis kinetics or experimental persistence.

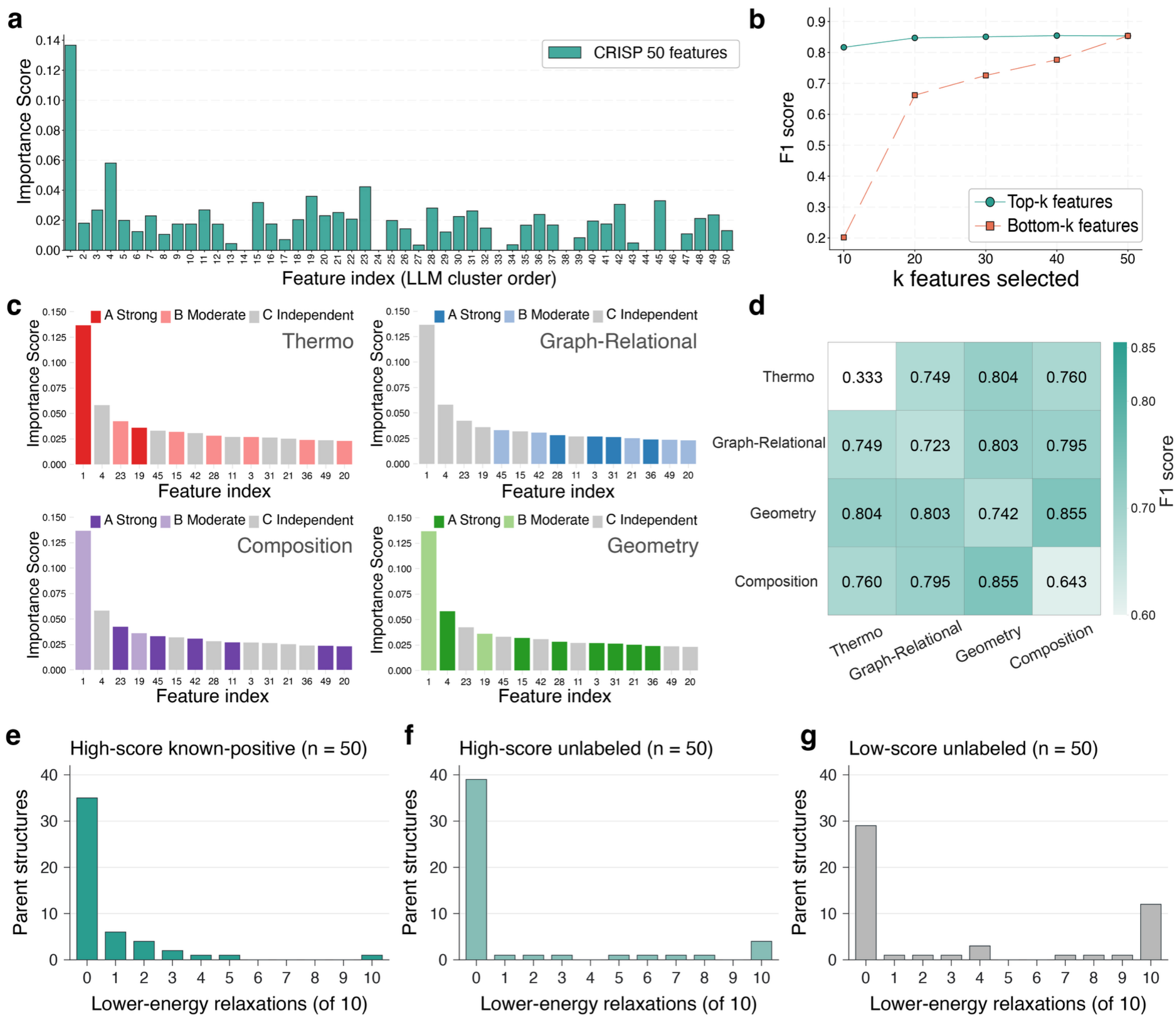

**Figure 4. Chemical interpretation and physical audit of CRISP predictions.** (a) Global impurity-based Random Forest importance of the 50 executable descriptors, indexed according to LLM rule-cluster frequency. (b) Estimated α-conditional PU F1 of models retrained using progressively larger subsets of the top- or bottom-ranked descriptors. Descriptor ranking was determined from the training data only. (c) Importance profiles of the leading descriptors within thermodynamic, graph-relational, geometric, and compositional families. Colors denote post hoc strong (A), moderate (B), and weak/independent (C) relevance assignments. (d) Estimated α-conditional PU F1 obtained using the A-rated descriptors from individual families and their pairwise unions; diagonal cells represent single-family models. (e–g) Parent-level distributions of the number of perturbations, among ten trials per structure, that relaxed to energies more than 0.02 eV/atom below the corresponding parent for high-scoring known-positive, high-scoring unlabeled, and low-scoring unlabeled structures, respectively. Each group contains 50 independent parent structures. Fewer energy-lowering perturbations are consistent with greater local structural robustness under the applied DFT protocol.

## Extending executable knowledge compilation to continuous materials properties prediction

Synthesizability is CRISP's primary test bed because it is a multifactorial outcome rooted in chemical heuristics, but the compilation workflow is not limited to classification. We therefore applied it to formation energy, a large and well-defined benchmark, and room-temperature ionic conductivity, a smaller and heterogeneous transport problem. Each task used a target-specific catalog compiled through the same workflow.

Formation-energy regression provided a data-rich test. CRISP achieved an MAE of 0.130 eV/atom and $R^2$ = 0.961 (**Figure 5a**). Under the same Random Forest consumer and splits, its MAE matched the expert-curated 17-descriptor representation (0.131 eV/atom) and improved on CFID (0.160 eV/atom), CrystalNN (0.344 eV/atom), Raw-CIF (0.387 eV/atom), and Robocrys (0.384 eV/atom; **Figure 5b**). The expert control was selected from a fixed candidate pool using chemical reasoning (**Table S8**), so the close agreement indicates that automatic compilation recovered much of the manually encoded information.

CRISP's advantage widened under the chemical-family shift. It retained the highest macro-averaged $R^2$ of 0.564, compared with 0.474 for the expert-curated representation and 0.323 for CFID; the remaining representations did not retain positive macro-averaged $R^2$ across the eight family holdouts (**Figure 5c**). The broader catalog therefore retained complementary information when an entire material family was excluded from training. Additional learning-curve, family-resolved, catalog-size, and expert-dimension analyses are provided in **Figures S17** and **S18**.

We next considered room-temperature ionic conductivity, a data-limited property governed by carrier concentration, site disorder, migration-network connectivity, and framework chemistry. On the fixed reduced-formula-group-disjoint split of 503 solid electrolytes, CRISP achieved a test MAE of 1.229 log units and $R^2$ = 0.570 (**Figure 5d**). Under the same split and learner, CRISP had the lowest MAE among the generic crystal-derived representations, followed by CFID at 1.329, Robocrys at 1.461, CrystalNN at 1.518, and Raw-CIF at 1.550 log units (**Figure 5e**).

We then fitted CRISP to all 503 development structures and evaluated it without refitting on 150 manually curated solid electrolytes absent from the reported dataset[29]. CRISP retained an MAE of 1.170 log units, $R^2$ = 0.476, and Spearman $\rho$ = 0.721 (**Figure 5f**), showing that the descriptors remained informative beyond the development collection. An 18-descriptor control based on manually selected ion-transport priors performed better. Because it embeds strong task-specific priors in a dataset of this size, we treat it as a specialized expert reference; its internal, reduced-data, and external results are reported in **Figure S20**.

The two regression tasks test CRISP in different data regimes. The formation-energy catalog matched the compact expert representation in distribution and retained the strongest signal across chemical-family shifts, whereas the ionic-conductivity catalog outperformed the generic representations and transferred to an independently curated cohort. Shear-modulus regression defines a complementary scope boundary, with fine-grained continuous geometric descriptors outperforming CRISP (**Figure S21**). Its utility is therefore task-dependent, but the same workflow can produce competitive and transferable representations for distinct continuous targets.

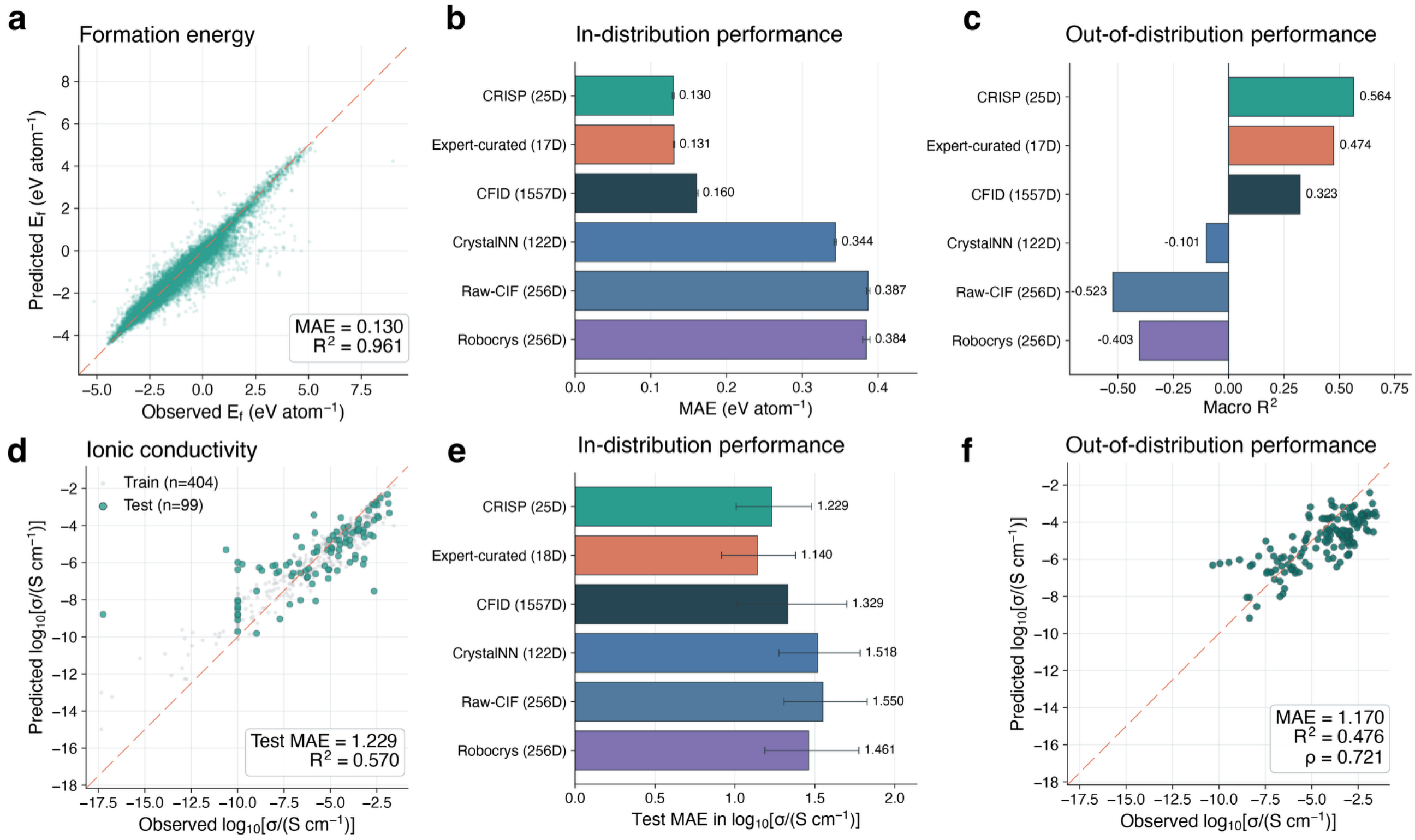

**Figure 5. Extension of CRISP to continuous materials-property prediction.** (a–c) Formation-energy regression. (a) Pooled held-out CRISP predictions across three material-identifier-grouped splits. The dashed line denotes ideal agreement. (b) Test MAE of CRISP, an expert-curated 17-descriptor representation, and four generic crystal-derived representations under the same Random Forest consumer and splits. Error bars show one standard deviation across the three splits. (c) Macro-averaged $R^2$ across eight held-out material families. (d–f) Room-temperature ionic-conductivity regression. (d) CRISP train and test predictions on the fixed split of 503 solid electrolytes. (e) Test MAE of CRISP, an expert-curated 18-descriptor representation, and four generic crystal-derived representations under the identical 404/99 split and Random Forest consumer. Error bars show 95% reduced-formula-group-clustered bootstrap intervals for the 99-structure test set. (f) CRISP predictions for an independently assembled 150-structure cohort after fitting to all 503 development structures, without refitting on the external cohort. Formation energy is reported in eV/atom and conductivity as $\log_{10}[\sigma/(S/cm)]$. Lower MAE and higher $R^2$ indicate better performance.

## Discussion

We introduced CRISP, a framework that formulates scientific representation construction as a rule-space exploration and compilation problem. CRISP systematically samples target-relevant chemical rules, consolidates related concepts, and compiles each into an executable descriptor without access to structures, labels, target values or data splits during representation construction. The resulting catalog is therefore a frozen, dataset-blind computational prior rather than a representation optimized against a particular benchmark.

For inorganic-crystal synthesizability, CRISP outperformed expert-curated and generic structural representations under a shared learner and retained the strongest predictive signal under structural-size and chemical-family shifts. Its transfer advantage was not confined to the rules most readily generated by the LLM: less frequently articulated concepts and descriptors that appeared partly redundant in distribution provided complementary information beyond the development domain. The DFT perturbation–relaxation audit further provided a physical consistency check by associating high CRISP scores with greater local structural robustness under the applied protocol. Across continuous-property tasks, CRISP matched the expert representation for formation energy in distribution, generalized most strongly across held-out chemical families, and retained predictive signal for ionic conductivity on an independently curated cohort. The stronger performance of task-specific expert descriptors for ionic conductivity and of finer continuous geometric representations for shear modulus shows that the value of compiled knowledge depends on the target and the resolution of information it requires.

The principal contribution of CRISP lies neither in using an LLM as another scientific predictor nor in generating interpretable features alone, but in systematically exploring and compiling a broad, stochastic space of chemical knowledge into reusable and inspectable programs. By making rule-space coverage an experimental variable, CRISP enables evaluation of when broadly compiled scientific priors improve prediction, particularly under distribution shift and limited data. Once compiled, the catalog supports prediction without per-material LLM inference or task-specific LLM fine-tuning, while exposing rule-level evidence to downstream explainers and screening systems (**Figure S22**). Although the present implementation used proprietary language models, the released rule texts, descriptor programs and audit records preserve the exact representation used for all reported results. More broadly, executable knowledge compilation provides a practical interface between language-based scientific knowledge and numerical learning systems while preserving the provenance and auditability of the resulting representation.

## Methods

### LLM-assisted discovery and consolidation of chemistry rules

CRISP was constructed by separating LLM-assisted chemistry-rule generation and compilation from statistical model training. For synthesizability, the rule-generation prompt instructed *gpt-4.1-mini-2025-04-14* to act as an expert in inorganic chemistry and propose ten rules that could indicate the synthesizability of an inorganic material from crystal information, with each explanation limited to 100 words. The request was submitted 1,000 times, producing 10,000 candidate natural-language rules. Repeated stochastic generation was used to reduce dependence on any single response and to sample both frequently stated principles and less common hypotheses.

The rule texts were embedded with *text-embedding-3-large*. Embedding vectors were L2-normalized, reduced by PCA, and normalized again before clustering. For the synthesizability catalog, k-means clustering used 50 clusters, 50 centroid initializations, a maximum of 1,000 iterations, and random seed 42. The constituent rule texts in each cluster were supplied to GPT-5 to synthesize one representative paragraph describing their shared chemical meaning. The prompts and procedural details are provided in **Supplementary Note 1**.

Formation-energy, shear-modulus, and ionic-conductivity rule discovery followed the same 1,000-call protocol with target-specific prompts. The LLM stages were dataset blind: no crystal structures, target values, data splits, model predictions, or performance metrics were provided. Each candidate rule was converted to a chemistry-focused keyword representation with *gpt-4.1-mini-2025-04-14*, embedded with *text-embedding-3-large*, reduced by PCA, and clustered into 25 rule families. GPT-5 synthesized one representative rule from each family. The changes between tasks were limited to the target-specific rule-discovery objective and catalog size.

### Conversion of natural-language rules into executable descriptors

Each consolidated synthesizability rule was converted into a Python descriptor function with *gpt-4.1-2025-04-14* at temperature 0. The prompt required one scalar floating-point output obtained through an explicit calculation. (**Supplementary Note 1**) Functions received a CIF-derived structure and, when required by the rule, fixed per-material physicochemical fields defined before model fitting. Standard inorganic-chemistry and structure-analysis libraries were allowed, whereas graph neural networks, pretrained synthesizability models, and target-trained surrogate predictors were prohibited. The 50 functions were assembled in fixed rule order.

The formation-energy, shear-modulus, and ion-conductivity catalogs used the same code-generation model and scalar-function requirement. Their prompts prohibited access to the corresponding targets or direct proxies, including material-specific formation or total energies, energy above hull, elastic tensors, elastic constants, moduli, stress–strain responses, database target retrieval, and target-trained surrogates. Static audits checked syntax, imports, forbidden tokens, and target leakage before execution. All 25 functions in each final catalog passed these gates.

Human intervention was limited to repairing syntax, type, import, and runtime errors while preserving the chemical calculation proposed by the model. The complete rule definitions, source code, prompt templates, and audit records are provided in the Supporting Information and GitHub repository.

### Crystal representations used for controlled comparison

The controlled synthesizability comparison evaluated six frozen representations on the same ordered material identifiers. CRISP comprised 50 executable chemistry-rule descriptors calculated from CIF-derived structural quantities, tabulated elemental attributes, and the fixed per-material physicochemical inputs required by individual rules. The expert-curated control contained 18 descriptors selected by a domain expert from a frozen 71-feature composition and CIF-derived candidate library which originated from pymatgen. Selection followed a written chemical rationale and was completed before inspecting material-level labels, CRISP results, or comparative benchmark performance.

Robocrystallographer (Robocrys) 256D embedded automated structure descriptions obtained from the Materials Project service when available or generated locally from the canonical CIF. CrystalNN 122D represented local coordination environments, CFID 1557D combined chemistry, cell, radial, angular, dihedral, and charge-related distributions, and Raw-CIF 256D embedded the complete UTF-8 CIF text with *text-embedding-3-large*. No synthesizability labels were used during featurization. Input provenance, dimensions, and candidate-time availability are reported in **Table S3**.

Regression tasks used the same four generic crystal-derived representations together with their task-specific CRISP catalogs. Expert-curated controls were selected before benchmark evaluation from the same frozen 71-feature library. The larger expert-ranked sets were retained only as dimensionality-sensitivity controls matching with 25D of CRISP.

### Development chronology and data separation

The 1,000-call budget, 50-rule synthesizability catalog, executable functions, and original MP30 split were established during the initial CRISP development. The same split was retained for the matched six-representation comparison to preserve direct comparability with the original benchmark. All new representation matrices and expert controls were frozen before comparative evaluation, and preprocessing, threshold calibration, feature ranking, and nested learning-curve subsets were fitted using training data only. The catalog-size analysis in **Figure S4** and the classifier comparison in **Figure S3** are retrospective sensitivity analyses rather than prospective model-selection experiments. Code repair was completed before benchmark evaluation and was restricted to execution failures as described above.

### Materials Project benchmark and positive–unlabeled labels

The primary synthesizability benchmark was constructed from a frozen Materials Project-derived dataset assembled in March 2024. MP30 was restricted to structures containing at most 30 atoms per unit cell representation and to entries compatible with the common structure-language benchmark. Entries whose generated structure descriptions contained 10,000 or more characters were excluded to maintain compatibility with the StructLLM pipeline.

A 90:10 split produced 90,174 training structures (34,512 P and 55,662 U) and 10,020 held-out test structures (3,835 P and 6,185 U). Material identifiers in this split were frozen and reused for CRISP, the representation comparison, compatible end-to-end baselines, classifier comparisons, and feature-subset analyses.

### Bagged positive–unlabeled random forest

For synthesizability, CRISP used a bagged positive–unlabeled random forest. In each of 50 bags, all positive training structures were paired with an equal-sized random subset of unlabeled structures. A RandomForestClassifier with 100 estimators and otherwise default scikit-learn settings was trained on each balanced bag. The final synthesizability score was the mean positive-class probability across the 50 forests.

For classifier selection, the random forest was compared with gradient-boosted trees, a multilayer perceptron, and logistic regression using the same 50-dimensional matrix, data split, and PU sampling protocol. The random forest was retained because it provided the strongest overall performance together with stable ensemble predictions and direct feature-level interpretation (**Figure S3**).

### Positive–unlabeled calibration and performance metrics

The primary synthesizability results use an α-calibrated operating point. Following previous PU synthesizability studies[5,9,11,12], $\alpha$ was fixed at 0.0567 and interpreted as the assumed prevalence of latent positives within U.

### Synthesizability baselines and distribution-shift evaluations

CRISP was compared with newly trained implementations of PU-CGCNN, CSLLM, StructLLM, and SynCry using the same frozen MP30 training and test identifiers. For each baseline, we reconstructed the input representation, model architecture, and training procedure described in the original study and then trained or fine-tuned the model on MP30, as appropriate. Thus, the comparison does not rely on previously published predictions or models trained on different data. All resulting score vectors were evaluated on the identical held-out identifiers using the same α-conditional PU metric definitions. This complete-pipeline comparison is distinct from the controlled six-representation experiment, in which the downstream PU-RF model, split, calibration, and evaluation protocol were held fixed.

For controlled size transfer, MP30-trained representation models were applied without retraining to 23,525 structures containing 31–49 sites. Chemistry transfer was evaluated by leaving out one of eight material families, retraining on the remaining families, and testing on the excluded family. Family-level results were macro-averaged, and minimum-to-maximum ranges were retained for visualization. For the paired CRISP–expert comparison, the eight prespecified family-level F1 values were treated as paired observations and evaluated with a two-sided exact sign test. The family split measures taxonomy-level transfer and is not element disjoint. Where repeated fits were performed, the inner fit/calibration partition and model sampling varied while the frozen outer test cohort was preserved. The temporal checkpoint compared an October 2022 Materials Project snapshot with positive evidence available by April 2026.

### Feature attribution and descriptor-family analyses

Global Random Forest importance was calculated by averaging impurity-based importance across the 50 forests. Sample-level TreeSHAP values were calculated separately for each forest and averaged across bags for the positive class. Additivity was checked by reconstructing ensemble probabilities from the averaged SHAP values and expected values. Repeated fits and held-out permutation importance were used to assess the stability of the leading descriptors.

For feature-subset evaluation, descriptor ranking was calculated from training bags only and then fixed before held-out evaluation. New 50-bag models were trained using progressively larger top- and bottom-ranked subsets. GPT-5 proposed initial A/B/C relevance labels, after which two domain experts reviewed the descriptors jointly and determined the final labels by consensus. Because independent ratings were not collected, inter-rater agreement is not defined. Models were retrained using the A-rated descriptors from individual physicochemical families and selected family unions.

### Regression targets and evaluation

Formation energy per atom and elasticity metadata were retrieved through mp-api 0.46.0 for the canonical MP30 identifiers. Legacy identifiers were resolved to current Materials Project identifiers before target joining, and target values were joined only after descriptor generation. Formation-energy regression retained 99,186 structures with finite formation energy per atom. Shear-modulus regression retained 10,794 structures with finite positive Voigt, Reuss, and Voigt–Reuss–Hill shear moduli that passed the consistency criteria reported in the Supporting Information.

The regression representations were evaluated using RandomForestRegressor models with 100 trees and otherwise default scikit-learn settings. Three random GroupShuffleSplit partitions assigned 20% of the structures to testing and used current material identifier as the grouping key. Training fractions of 5, 10, 20, 40, 60, 80, and 100% were nested within each split and shared across representations. Median imputation and zero-variance removal were fitted on training data only. Metrics included MAE, RMSE, $R^2$, and Spearman $\rho$. Chemistry transfer used the same eight held-out material families as the synthesizability analysis, with three model seeds per family.

The ionic-conductivity development dataset contained 503 inorganic solid electrolytes with room-temperature

$\log_{10}[\sigma/(S/cm)]$ targets. The fixed reduced-formula-group-disjoint split assigned 404 structures to training and 99 to testing. These data were retrieved from a previous study[29]. External validation used a manually curated cohort of 150 solid electrolytes whose structures were absent from the previously reported 503-structure development collection.

The generic representation comparison included CRISP 25D, CrystalNN 122D, CFID 1557D, Raw-CIF 256D, and Robocrys 256D on the same 503 structures, split, and Random Forest consumer. Raw-CIF and Robocrys embeddings were generated with *text-embedding-3-large*. No representation-specific hyperparameter tuning was performed.

Reduced-data sensitivity was evaluated using 20 shared nested selections of reduced-formula groups at 10, 25, 50, 75, and 100% of the training partition. Preprocessing was refitted within each subset, and the 99-structure test set remained unchanged.

### DFT perturbation–relaxation analysis

Parent structures were DFT-relaxed geometries obtained from the Materials Project. Each of the High-score known-positive, High-score U, and Low-score U groups contained 50 parent structures. Ten perturbed structures were generated per parent by randomly displacing all fractional coordinates by 1%, with the same ten random seeds applied across groups. Each perturbed structure was relaxed using settings consistent with its parent and followed by a static single-point calculation; parent energies were obtained using the same static settings. Energy changes were normalized per atom relative to the corresponding parent and evaluated at –0.01, –0.02, and –0.03 eV/atom thresholds. The analysis is exploratory and tests association with local structural robustness under this perturbation–relaxation protocol; it does not identify kinetic trapping or experimental realizability.

DFT calculations were performed with VASP 5.4.4[30], the projector-augmented-wave method[31,32], and the PBE[33] exchange-correlation functional. Spin-polarized calculations used a 520 eV plane-wave cutoff. Atomic positions and lattice parameters were relaxed, followed by static calculations. Electronic convergence criteria were $1 \times 10^{-4}$ eV for relaxation and $1 \times 10^{-5}$ eV for static calculations; ionic relaxation used an energy-change criterion of $1 \times 10^{-3}$ eV. Gaussian smearing of 0.05 eV and Γ-centered k-point meshes were used. Structural similarity was quantified by Euclidean distance between CrystalNN-based fingerprint vectors implemented in matminer.

## Supporting Information

The Supporting Information contains task-specific prompts, rule-consolidation and code-audit procedures, executable descriptor catalogs, and representation provenance; descriptor-attribution, family-resolved, stability-regime, and DFT analyses; and formation-energy, ionic-conductivity, and shear-modulus regression controls.

## Data availability

The data supporting this study are provided in the Article, Supplementary Information, and accompanying Supporting Data. Crystal structures and associated metadata used for the MP30 analyses are publicly available from the Materials Project. The 503-structure ionic-conductivity development dataset is available through Zenodo at https://doi.org/10.5281/zenodo.17157647.

## Code availability

The source code used for rule consolidation, executable-descriptor generation, model training, evaluation and figure preparation, together with the prompts, frozen rule catalogs and executable descriptor functions, is available at https://github.com/snu-micc/CRISP.

## Acknowledgements

This work was supported by Hyundai Motor Group.

## Author contributions

J.C. conceived the study, developed the methodology and software, performed the analyses and validation, curated the data, generated the figures, and wrote the original manuscript. K.J. performed the DFT perturbation–relaxation analysis, curated and validated the ionic-conductivity datasets, and contributed to data interpretation, scientific discussion and manuscript writing. S.K. and S.C. contributed to the initial development of the research concept and subsequent discussions. K.N., S.H.N. and D.K. contributed to project discussions and reviewed the manuscript. Y.J. supervised the project, contributed to conceptualization and interpretation, acquired funding, and revised the manuscript. All authors reviewed and approved the final manuscript.

## Competing interests

The authors declare the following competing interests: this work was supported by Hyundai Motor Group. K.N., S.H.N. and D.K. are employees of Hyundai Motor Company. The remaining authors declare no competing interests.